\documentclass[12pt,a4paper]{article}
\usepackage{graphicx}

\usepackage{amsmath,amssymb}
\usepackage{epsfig}
\usepackage{color}
\numberwithin{equation}{section} 
\newcommand{\Frac}[2]{\frac{\displaystyle #1}{\displaystyle #2}}

\def\cV{{\cal V}}
\def\cG{{\cal G}}
\def\cI{{\cal I}}
\def\cN{{\cal N}}
\def\cD{{\cal D}}
\def\cI{{\cal I}}

\def\Dslash{\not{\hbox{\kern-4pt $D$}}}
\def\dslash{\kern-4pt \not{\hbox{\kern-2pt $\partial$}}}
\def\pslash{\not{\hbox{\kern-2pt p}}}

\begin{document}

\providecommand{\V}{\mathcal{V}} \providecommand{\rh}{\hat{r}}

\begin{titlepage}

\begin{center}
\renewcommand{\baselinestretch}{1.8}\normalsize
\textbf{\LARGE On the Evaluation of Gluon Condensate Effects  in the
Holographic Approach to QCD}
\\[1.5cm]

{\large Luigi~Cappiello$\, ^{1,2}$ and Giancarlo~D'Ambrosio$\, ^2$}\\[0.5cm]

\renewcommand{\baselinestretch}{1.2}\normalsize
\emph{$^1$ Dipartimento di Scienze Fisiche, Universit\`{a} di
Napoli
``Federico II''}\\
{and}\\
\emph{$^2$ INFN -- Sezione di Napoli,}\\[10pt]
\emph{Via Cintia, 80126 Napoli, Italy}
\end{center}

\begin{abstract}
In holographic QCD  the effects of gluonic condensate  can be
encoded in a suitable deformation of the 5D metric. We develop two
different methods for the  evaluation of first order perturbative
corrections to masses and decay constants of vector resonances  in
5D Hard-Wall models of QCD due to small deformations of the
metric. They are extracted either from a novel compact form  for
the first order correction to the vector two-point function, or
from perturbation theory for vector bound-state eigenfunctions:
the equivalence of the two methods is shown. Our procedures are
then applied to flat and to AdS 5D Hard-Wall models; we complement
results of  existing literature evaluating the corrections to
vector  decay constant and to two-pion-one-vector couplings: this is particularly relevant to satisfy
the sum rules. We
concentrate our attention on the effects for the Gasser-Leutwyler
coefficients; we show that, as in the Chiral Quark model, the
addition of the gluonic condensate improves the consistency, the understanding
and  the agreement with phenomenology of the
holographic model.
\end{abstract}
\vfill \emph{E-mails:}
\texttt{Luigi(dot)Cappiello,Giancarlo(dot)Dambrosio(at)na(dot)infn(dot)it}
\end{titlepage}

\section{Introduction}
The original Maldacena conjecture \cite{Maldacena:1997re} on the
AdS/CFT duality between ${\cal N}=4~SYM$ and string theory on
$AdS_5 \times S^5$ has been extended to a conjectured holographic
equivalence between four-dimensional (4D) strongly coupled QCD at
large $N_c$ and a 5D weakly interacting gauge theory coupled to
gravity on a 5D space $M_5$  not necessarily (even asymptotically)
$AdS_5$. New ingredients have to be introduced in order to comply
with the effective low energy description of a confining theory
such as QCD. One possibility is to modify the background geometry
of the dual gravity theory,  as in the Sakai-Sugimoto model
\cite{Sakai:2004cn}, or, as in the phenomenological Hard-Wall (HW)
models  of \cite{Erlich:2005qh} and \cite{Hirn:2005nr}, one can
cut-off the $AdS_5$ at a finite size, producing  confinement and
an infinite spectrum of Kaluza-Klein resonances. The two cited HW
models differ in the implementation of  spontaneous chiral
symmetry breaking ($\chi SB$), another fundamental ingredient of
low energy hadron dynamics. In \cite{Erlich:2005qh}, $\chi SB$ is
induced by a 5D scalar field, having the right properties to be
 the dual of the $\bar q q$ operator of QCD, whose
non-vanishing vacuum expectation value (VEV) is responsible for
$\chi SB$. In the approach of \cite{Hirn:2005nr}, $\chi SB$
follows from imposing appropriate infrared (IR) boundary
conditions.

The Soft-Wall (SW) model, proposed in \cite{Karch:2006pv}, is
instead a 5D holographic model in which the $AdS_5$ has no
cut-off, but confinement and an  infinite spectrum of Kaluza-Klein
resonances are due  a non-trivial dilaton background. The
phenomenological advantage of the SW model is the fact that,
contrary to the what happens for the KK states of the HW models,
the spectrum of resonances follows a linear Regge-trajectory,
which leads to a better agreement with the corresponding QCD
resonance spectrum in the intermediate energy regime.

In  any of  models above,  one boldly conjectures the
applicability of the following holographic recipe for the
calculation of correlation functions of the dual 4D theory.  For
every quantum operator ${\cal O}(x)$ in QCD, there exists a
corresponding 5D field $\phi(x,z)$, fixed  by the boundary
condition $\phi(x,0) \equiv \phi_0(x) $ on the ultraviolet (UV)
boundary of $M_5$. The generating functional of the correlation
functions of the 4D theory, is equal to  the 5D  action evaluated
\emph{on-shell}:
\begin{align}
\exp \left(W_4[\phi_0(x)]\right)\equiv\langle \mbox{exp}(i\int d^4
x \,\phi_0(x)\,{\cal O}(x))\rangle_{\rm QCD_4} =
\mbox{exp}\left(iS_{\rm M_5}[\phi_0(x)]\right) \label{holography}.
\end{align}
%

Varying $S_{\rm M_5}[\phi_0(x)]$   with respect to the sources
$\phi_0(x)$, and then setting them to zero, one gets the connected
$n$-point Green's functions of the 4D theory.

Having a 5D  holographic  description of low energy QCD makes
compelling the comparison with 4D low energy QCD models: VMD,
large $N_C$,  Chiral Quark Model, ... QCD at low energy is
described by chiral perturbation theory and thus the question
amounts to how good is the prediction for the Gasser-Leutwyler
coefficients, $L_i$. It is fair to say that all these models, 4D
and 5D, compare fairly well with the $L_i$'s phenomenological
values: a more specific question is if the presence of gluonic
condensate improves the agreement with phenomenology. Actually
this question has a very neat answer in the Chiral Quark Model
\cite{Manohar:1983md, Espriu:1989ff}: the agreement with
phenomenology improves \cite{Espriu:1989ff}. So we think that this
is a very well motivated question for holographic QCD.

In the holographic dictionary, 4D condensates,  \emph{i.e.}
non-vanishing VEV's of some  composite operators of the 4D theory,
are  related to the behavior near the UV boundary  of the dual 5D
fields \cite{Klebanov:1999tb}. The effect of a gluonic condensate
can be induced by a non-trivial dilaton field, or, equivalently,
by a suitable deformation of the original $AdS_5$ metric, as we
will discuss in detail.

We shall be concerned with the  correction to the two-point
functions  in presence of small deformations of the 5D metric
background. Here small means that we can treat them as
perturbations and we shall calculate the first order effect on the
two-point function of the theory. Although applicable also  to the
evaluation of  two-point functions of other 4D operators, we shall
consider only the case of conserved (flavor) currents, $ \langle
J_\nu (x) J_\mu (y) \rangle$. So, instead of  generic 5D actions,
we shall consider only the case of HW models with 5D vector gauge
fields, as they are the fields dual to these currents.

After a  a short review  of the holographic prescription in Sec.
2,  we compute   in Sec.3 the  first order correction to the
vector  two-point function and  show how to cast them in a very
compact form. With hindsight, we justify the result deriving it
directly from the holographic prescription. We  present also
another approach to evaluate directly  how the metric deformation
affects masses and  decay constants of the vector resonances,
which appear as intermediate states in the bound-state expansion
of the two-point function. We do it by developing  perturbation
theory for the resonance wave-functions. The resulting formulae
differ from those of usual  perturbation theory, since a
deformation of the metric affects also the scalar product of
wave-functions. The interest of this approach lies not only in the
fact that it provides a different route to the evaluation of the
perturbation of resonance masses and decay constants, which could
be alternatively extracted from the behavior of the perturbed
two-point around its singularities: in fact, since  it directly
provides the perturbed wave-functions, it allows, for instance,
the evaluation of the effects of the metric deformation on other
low energy coupling such as the two-pion-one-vector coupling. The
formal equivalence of the two methods is shown in the Appendix.

 In Sect.4, we consider first the case of an  HW model with flat 5D metric. It is
computationally simpler and we use it to illustrate our methods in the evaluation of the corrections
to resonance masses and decay constants: these results are original.
We then examine the
case of HW model where the extra-dimension is an $AdS_5$ slice
\cite{Erlich:2005qh}, \cite{Hirn:2005nr}. We
also consider some effects on  the axial vector resonance, for
which we adopt the description in \cite{Hirn:2005nr}. Moreover, perturbation
theory for the resonance wave-function allows us to evaluate also
the corrections to the two-pion-one-vector couplings. We want to
stress that our results represent a novelty, since no expressions
for the corrections to the vector decay constants and to the
two-pion-one-vector couplings have appeared in the literature even in
papers explicitly devoted to this effect in the $AdS_5$ HW model.
This was one of the motivations of our work and constitutes the core of our original analytic work.

 In
holographic models, the fact that all vector resonance wave
functions are Kaluza-Klein excitations of a 5D gauge field has the
consequence that they satisfy completeness relations from which
sum rules can be obtained. We address the issue of the saturation
of these sum rules by the first few resonances both in the
unperturbed and in the perturbed metric case. This relevant discussion is possible only after our novel calculation of the  corrections to the vector decay constants and to the
two-pion-one-vector couplings.

 In  Sec. 5   we present   numerical analyses and conclusions:  we evaluate  the effects  of the gluon condensate on low energy
parameters of QCD, in particular the Gasser-Leutwyler
coefficients, the pion decay constant and $\rho-$meson parameters.
Our discussion leads to a remarkable analogy between the  $AdS_5$ HW model and the
  Chiral Quark model.

\section{Two-point function of vector gauge fields in
Hard-Wall models}

We consider  a 5D  holographic Hard-Wall models  with the
extra-dimension $z$ restricted to a finite interval  $0\leq z\leq
z_0$ and  unperturbed metric written as a conformal factor times
the 5D flat metric as  follows:
\begin{align}
g_{MN}dx^Mdx^N = w(z)^2\left(\eta_{\mu \nu}dx^{\mu}dx^{\nu} -
dz^2\right) \label{metric} ,
\end{align}
where $ \eta_{\mu\nu} = {\rm Diag}\, (1,-1,-1,-1) $ and $ \mu, \nu
= (0,1,2,3) $, $ M, N = (0,1,2,3,z) $. The conformal factor $w(z)$
is left unspecified; in the case of $AdS_5$ it would be
$w_{AdS}(z)=1/z$. We consider $z=0$ as the UV-brane, where the
bulk fields have to coincide with the external sources of the 4D
theory and $z=z_0$ as the IR-brane where  suitable boundary
conditions are imposed. As we shall mainly consider vector gauge
fields, we impose  Neumann-type boundary conditions, which are
common to all the HW models mentioned in the introduction, where,
moreover, these b.c. force the absence of a massless resonance
mode.
 In the case of SW models, where  the fifth coordinate
 $z$ is no more restricted to a finite interval, the IR boundary
 condition  would be  replaced by a normalization condition.

 The quadratic part of the 5D action for the gauge fields is
given by:
\begin{equation}
\label{SAdS1} S_{M_5} = -\frac{1}{4g_5^2} \int d^4x \int_0^{z_0}
 dz\sqrt{g}\, g^{MN}g^{RS}F_{MR}F_{NS}
\end{equation}
with  linearized field strength $F_{MN}=\partial_M A_M-\partial_N
A_M$ and $g_5^2$ is a 5D coupling constant, which in $AdS_5$
models is usually fixed to $g_5^2=12\pi^2/N_c$, where $N_c$ is the
number of colors of the 4D dual gauge theory, by matching the
two-point function at large Euclidean momentum with the
perturbative result of QCD parton loop. Otherwise, $g_5^2$ could
be fixed by the requirement of obtaining the physical value of the
pion decay constant $f_{\pi}$.  Its actual value is irrelevant for
the general discussion of this and the next Sections, so we shall
momentarily suppress it.
 For the same reason, we  have also suppressed any flavor index of the
 5D  fields.
We shall restore the correct factors in Sec. 4.2, when we discuss
the 5D AdS slice, which is the playground of the HW models
\cite{Erlich:2005qh, Hirn:2005nr}. We shall need them in order to
make some numerical analysis of the effects of condensates and
compare with the existing literature.

It is convenient to work in $A_z = 0$ gauge. Then, the  5D gauge
fields $A_{\mu}(x,z)$   holographically correspond to  conserved
vector currents $J_{\mu}(x)$ of the dual 4D theory, and the
holographic formula will allow the calculation of the
 correlation function of two currents, in two steps.
First, one  solves the 5D equations of motion of the gauge field,
requiring  the solution to coincide on the UV boundary with the 4D
source $A_{\mu}(x)$ of the vector current. This is done by means
of  the bulk-to-boundary propagator.  Secondly,  the 5D action is
evaluated  on this solution and identified with the generating
functional of the 4D theory according to eq.(\ref{holography}).
Finally, by varying twice with respect to the boundary sources,
one obtains the holographic result for $<J_{\mu}(x)J_{\mu}(y)>$.

Using Fourier-transformed gauge fields, written as $\tilde
A_{\mu}(p,z) = \tilde{A}_{\mu}(p)\cV(p,z)$, where $\tilde
A_{\mu}(p,z)$ and $\tilde{A}_{\mu}(p)$ are the Fourier transforms
of $A_{\mu}(x,z)$ and the source $A_{\mu}(x)$ respectively, one
gets:
\begin{equation}
S_{{\rm AdS}_5}^{(2)} = -\int \frac{d^4p}{(2 \pi)^4}
\tilde{A}^{\mu}(p)\tilde{A}_{\mu}(p)\left( w(z)\partial_z
\cV(p,z)\right)|_{z=\epsilon} \label{S2q}.
\end{equation}
 The
bulk-to-boundary propagator, $\cV(p,z)$, satisfies the linearized
5D  equations of motion with 4D momentum, $p$:
\begin{equation}
\partial_z\left(w(z)\partial_z\cV(p,z)\right)+p^2w(z)\cV(p,z)=0
\label{EOM}
\end{equation}
and the boundary conditions
\begin{equation}
\cV(p,0)=1, \;\; \partial_z \cV(p,z_0)=0\label{BC0}
\end{equation}
Varying (\ref{S2q}) with respect to the boundary sources gives the
scalar part of the two-point function:
\begin{equation}
\Sigma(p^2) = -\left(w(z)
\partial_z \cV(p,z)\right)_{z = \epsilon \rightarrow 0} \label{sigma0}
\end{equation}
which is in general defined by
\begin{equation}
\int~d^4x \ e^{i p \cdot x}\langle J_{\mu}(x)J_{\nu}(0) \rangle =
\left( \eta_{\mu \nu} - \frac{{p_{\mu}p_{\nu}}}{p^2}
\right)\Sigma(p^2) \ .
\end{equation}
Let the functions $\varphi_n(z)$ form  the orthogonal basis of
normalizable eigenfunctions  satisfying
\begin{equation}
\partial_z\left(w(z)\partial_z\varphi_n(z)\right)+m_n^2w(z)\varphi_n(z)=0
\end{equation}
and the boundary conditions:
\begin{equation}
\varphi_n(0)=0, \;\; \partial_z \varphi_n(z_0)=0.\label{bceigen}
\end{equation}
They are normalized according to the scalar product
\begin{equation}
<\varphi_m,\varphi_n>\equiv
\int_0^{z_0}\varphi_m(z)\varphi_n(z)w(z)dz\label{scalarproduct}
\end{equation}
in order to get  from the 5D action a canonically normalized 4D
kinetic term for the Kaluza-Klein modes $A^{(n)}_{\mu}(x,z) =
A^{(n)}_{\mu}(x)\varphi_n(z)$. They also satisfy the completeness
relation
\begin{equation}
\sum_{n=1}^\infty w(z')\varphi_n(z)\varphi_n(z')=\delta(z-
z')\label{completeness}
\end{equation}
The two-point function $\Sigma(p^2)$ admits a bound-state
expansion given by
\begin{equation}
\Sigma(p^2)=\sum_{n=1}^\infty\frac{F^2_n}{p^2-m_n^2}\label{boundstates}
\end{equation}
with
\begin{equation}
F^2_n=\left(\left(w(z)
\partial_z \varphi_n(z)\right)_{z = \epsilon \rightarrow
0}\right)^2.
\end{equation}

Eq. (\ref{boundstates}) shows the pole structure of the two-point
function,  corresponding to an infinite sum over resonances which
 correspond  vector mesons with increasing masses $m_n$
and decay constants $F_n$.

It is also useful to introduce the Green function $\cG(p, z,z')$,
which is the solution of the equation
\begin{equation}
\partial_z\left(w(z)\partial_z\cG(p, z,z')\right)+p^2w(z)\cG(p,
z,z')=\delta(z-z')
\end{equation}
with the boundary conditions
\begin{equation}
\cG(p, 0,z')=0, \;\; \partial_z \cG(p, z_0,z')=0
\end{equation}
Then $\cG(p, z,z')$ can be expanded as
\begin{equation}
\cG(p,z,z')=\sum_{n=1}^\infty\frac{\varphi_n(z)\varphi_n(z')}{p^2-m_n^2}
\end{equation}
and the bulk-to-boundary propagator $\cV(p,z)$ and the two-point
function $\Sigma(p^2)$ can be respectively written as:
\begin{equation}
\cV(p,z)=\left(w(z')
\partial_{z'}\cG(p,z,z')\right)_{z' = \epsilon \rightarrow 0}
\end{equation}
and
\begin{equation}
\Sigma(p^2)=\left(w(z)\partial_{z}w(z')
\partial_{z'}\cG(p,z,z')\right)_{z =z' =\epsilon \rightarrow 0}
\end{equation}

\section{First order corrections  due to a metric perturbation}
\subsection{Correction to the two-point function}
Let us now suppose  the conformal factor in the metric
(\ref{metric}) to be  written as
\begin{equation}
w(z)=w_0(z)(1+h(z)),
\end{equation}
with $|h(z)|< 1$,  that can treated perturbatively, and $h(0)=0$,
so as  not to alter the behavior of the metric on the UV boundary
$z=0$. Then, writing the bulk-to-boundary as
 $\cV(p,z)=\cV_0(p,z)+\cV_1(p,z)$, one has that
the unperturbed $\cV_0(p,z)$ obviously satisfies (\ref{EOM}) with
$w(z)$ replaced by $w_0(z)$ and the boundary conditions
(\ref{BC0}), while  the first order
 correction $\cV_1(p,z)$  is a solution of
\begin{equation}
\partial_z\left(w_0(z)\partial_z\cV_1(p,z)\right)+p^2w_0(z)\cV_1(p,z)=-w_0(z)D_1\cV_0(p,z),
\end{equation}
where the differential operator $D_1$ is
\begin{equation}
D_1\equiv \partial_z h(z)\partial_z,\label{D1}
\end{equation}
 and satisfies homogeneous boundary conditions
\begin{equation}
\cV_1(p,0)=0, \;\; \partial_z \cV_1(p,z_0)=0.
\end{equation}
Let $f_1(p,z)$ and $f_2(p,z)$ be two independent solutions of the
differential equation for $\cV_0(p,z)$, then, imposing the
boundary conditions (\ref{BC0}) it is easy to get
\begin{equation}
\cV_0(p,z)=\frac{f'_2(p,z_0)f_1(p,z)-f'_1(p,z_0)f_2(p,z)}
{f'_2(p,z_0)f_1(p,\epsilon)-f'_1(p,z_0)f_2(p,\epsilon)},\label{V0}
\end{equation}
where prime denotes derivative with respect to $z$.  As we shall
see in the next Section, in the case of flat 5D slice $\cV_0(p,z)$
is expressed in terms of trigonometric functions, while in  the
case of $AdS_5$ slice, $w_0(z)=1/z$ and one gets for $\cV_0(p,z)$
the well known ratio of combinations of Bessel functions
\cite{Pomarol:1999ad, ArkaniHamed:2000ds}.

Standard methods, \emph{e.g.} the use of the Green function, lead
to the following expression for $\cV_1(p,z)$
\begin{eqnarray}
\cV_1(p,z)=\frac{f_2(p,\epsilon)f_1(p,z)-f_1(p,\epsilon)f_2(p,z)}
{f'_2(p,z_0)f_1(p,\epsilon)-f'_1(p,z_0)f_2(p,\epsilon)}\left(\cI_1
f'_2(z_0)-\cI_2f'_1(z_0)\right) \nonumber\\
+\frac{1}{c}\displaystyle\int_\epsilon^{z}
\left(f_2(p,z)f_1(p,\zeta)-f_1(p,z)f_2(p,\zeta)\right)g(\zeta)w_0(\zeta)d\zeta
\label{V1}
\end{eqnarray}
where
\begin{equation}
g(z)\equiv -h'(z)\cV'_0(p,z)=-D_1\cV_0(p,z),
\end{equation}
and
\begin{equation}
\cI_i\equiv\frac{1}{c}\displaystyle\int_\epsilon^{z_0}f_i(p,\zeta)g(\zeta)w_0(\zeta)d\zeta,
\;\;i=1,2 \end{equation}
 and $c=w_0(z)W(z)$,
where $W(z)=f'_2(p,z)f_1(p,z)-f'_1(p,z)f_2(p,z)$ is the Wronskian
of the two functions $f_1(p,z)$ and $f_2(p,z)$.

Previous formulae have been written with an UV cut-off
$z=\epsilon>0$ to comply with UV divergences which my appear in
explicit examples, \emph{e.g.} $AdS_5$, due to the divergent
behavior of the conformal factor $w_0(z)$ as $z\rightarrow 0$.

From the expressions for $\cV_0(p,z)$ and $\cV_1(p,z)$ one obtains
the first order corrected   two point function
\begin{equation}
\Sigma(p^2)=\Sigma_0(p^2)+\Sigma_1(p^2)= -\left(w_0(z) \partial_z
\cV_0(p,z)\right)_{z = \epsilon \rightarrow 0}-\left(w_0(z)
\partial_z \cV_1(p,z)\right)_{z = \epsilon \rightarrow 0}
 \label{sigmacorr}
\end{equation}
where we have assumed that $w_0(z)h(z)\rightarrow 0$ for
$z\rightarrow 0$ to get rid of an additional term.

Explicit use of the Green function method to get the first order
correction to the two-point function due to metric perturbation in
the $AdS_5$ case was made in \cite{Erlich:2006hq}, where it is also
mentioned that the method   had been used by the authors of
\cite{Hirn:2005vk}. The expressions in \cite{Erlich:2006hq} are
readily obtained from (\ref{V1}) with $f_1(p,z)$ and $f_2(p,z)$
being, in that case, Bessel functions. We shall deal with the
$AdS_5$ HW model in Sec. 4.2.

Although (\ref{sigmacorr}) gives the solution to the problem of
finding the first order correction to the two-point function, due
to metric perturbations,  it rests on the choice of two solutions
$f_1(p,z)$ and $f_2(p,z)$ of eq.(\ref{EOM}). It would be more
satisfying if the first order correction could be written in term
of the unperturbed bulk-to-boundary  propagator (\ref{V0}) and of
the perturbation operator $D_1$ (\ref{D1}). Indeed, it is the
case. Using the expression (\ref{V0}) for $\cV_0(p,z)$ and further
manipulating  the expression (\ref{sigmacorr}), the first order
correction to the two-point function can be written in the
following two equivalent forms obtained  one from the other by
integration by parts:
\begin{eqnarray}
\Sigma_1(p^2)=-<\cV_0(p,z), D_1 \cV_0(p,z)>_0=
\displaystyle\int_\epsilon^{z_0}w_0(z)h(z)\left((\cV'_0(p,z))^2-p^2\cV_0(p,z)^2\right)dz
\label{compact}
\end{eqnarray}
where $<f,g>_0$ denotes the scalar product (\ref{scalarproduct})
with respect to the unperturbed metric.

 Both expressions in (\ref{compact}) have interesting interpretations. The
 first shows  the  first order correction to the two-point
 function as the \lq\lq matrix element" of the perturbation operator (\ref{D1})
 on the unperturbed bulk-to-boundary propagator $\cV_0(p,z)$, with
 the scalar product defined by the unperturbed metric.
  The second expression in (\ref{compact})
 shows that it is given by the 5D  action on-shell, but with
 the perturbed conformal factor of the metric $w_1(z)=w_0(z)h(z)$.

The last  result suggests the existence of  a simpler derivation.
In fact, as we shall see in a moment, it is a direct consequence
of the AdS/CFT recipe (\ref{holography}) of identifying the 4D
effective action with the 5D  action evaluated on-shell.

In momentum space, the $z$-dependent part of the 5D bulk action
on-shell is given by
\begin{equation}
S_{M_5}[\cV]= -\displaystyle\int_{0}^{z_0}w(z) \left(\left(
\cV'(p,z)\right)^2-p^2\cV(p,z)^2\right)dz,
\end{equation}
where $\cV(p,z)$ is the bulk-to-boundary propagator of 5D vectors
in the metric background with warp factor $w(z)$.  If we write
$w(z)=w_0(z)(1+h(z))$ and look for a perturbative solution of the
form $\cV(p,z)=\cV_0(p,z)+\cV_1(p,z)$ we have  at first order:
\begin{eqnarray}
&\hspace{-2cm}S_{M_5}[\cV]=S_{M_5}^{(0)}[\cV_0]+<\left[\Frac{\delta
S_{M_5}^{(0)}}{\delta
\cV}\right]_{\cV_0}, \cV_1(p,z)>_0\label{Sexpa}\\
&-\displaystyle\int_{0}^{z_0}w_0(z)h(z) \left(\left(
\cV'_0(p,z)\right)^2-p^2\cV_0(p,z)^2\right)dz,\nonumber
\end{eqnarray}
On-shell,  the first term becomes a boundary term, i.e. the 0-th
order term in (\ref{sigmacorr}), while the second term,
proportional to the unperturbed equation of motion, vanishes and
we are left with just the last one, which is precisely what we got
in (\ref{compact}).

As we have already mentioned,  one can use (\ref{compact}), for
instance, to evaluate sub-leading corrections to the asymptotic
expansion of $\Sigma(Q^2)$ for large Euclidean momentum
$Q^2=-p^2$. More interestingly, one can use it in order to extract
corrections to the parameters, masses and decay constants,
appearing in the bound-state decomposition of the two-point
function, by studying the structure of pole singularities of the
two-point function.  In fact, the correction alters both the
position of the poles and the corresponding residues. There are,
however, some little subtleties in this procedure which we shall
illustrate in Sec.4.

\subsection{Corrections to the resonance wave functions}

The two-point function can be written as the bound-state
decomposition (\ref{boundstates}), where masses and decay
constants are  related to the mass eigenvalues and to the values,
weighted by the metric conformal factor, of the derivatives of the
wave functions $\varphi_n(z)$ at the origin. In presence of a
deformation of the metric,  wave functions are modified too. First
order corrections can be evaluated using perturbation theory for
eigenfunction. The derivation is straightforward, but leads to
final formulae  slightly different from those familiar from
quantum mechanics. The reason is that the deformation of the
metric does not only affect the wave-function equations, but also
the scalar product entering in the normalization condition
(\ref{scalarproduct}). This leads to a new diagonal term in the
first order correction of $\varphi_n(z)$, which is normally absent
in perturbation theory.

If  $\lambda^{(0)}_n\equiv -(m_n^{(0)})^2$ and
$\varphi_n^{(0)}(z)$ denote the eigenvalues and the
eigenfunctions of the unperturbed case $w(z)=w_0(z)$, then  the
first order corrections
$\lambda_n=\lambda^{(0)}_n+\lambda^{(1)}_n$ and $\varphi_{
n}(z)=\varphi_{ n}^{(0)}(z)+\varphi_{ n}^{(1)}(z)$ are given by
the following expressions
\begin{eqnarray}
\lambda^{(1)}_n&=&<\varphi_{ n}^{(0)}, D_{1}\; \varphi_{
n}^{(0)}>_0
\label{correlambda}\\
\varphi_{ n}^{(1)}(z)&=&-\frac{1}{2}< \varphi_{ n}^{(0)},
\varphi_{ n}^{(0)}h>_0 \varphi_{ n}^{(0)}(z)\nonumber\\
&+&\displaystyle\sum_{m\neq n}\frac{< \varphi_{ m}^{(0)}, D_{1}\;
\varphi_{ n}^{(0)}>_0}{\lambda_m^{(0)}-\lambda_n^{(0)}}
\varphi_{m}^{(0)}(z),\label{correeigen}
\end{eqnarray}
where $D_1$ is defined in (\ref{D1}). The first term in
(\ref{correeigen}) is the one due to the fact that the deformation
of the metric affects also the scalar product. In explicit
examples, it is numerically  relevant.

From (\ref{correlambda}) and (\ref{correeigen}) one gets the
 the corrections to masses and coupling constants:
\begin{eqnarray}
m_n&=&m_n^{(0)}\left(1-\Frac{1}{2}\Frac{<\varphi_{ n}^{(0)},
D_{1}\;
\varphi_{ n}^{(0)}>_0}{(m_n^{(0)})^2}\right)\label{pertmass}\\
F_n &=&F_n^{(0)}\left(1+\Frac{(\varphi_{ n}^{(1)})'(z)}{(\varphi_{
n}^{(0)})'(z)}\right)_{z=\epsilon\rightarrow 0}\label{pertdecay}
\end{eqnarray}

It may seem not evident  that the corrections to masses
 and decay constants obtained here generate,
through the bound-state decomposition (\ref{boundstates}), the
same correction to the two-point (\ref{compact}) deduced in the
previous Section, but this is the case, as explicitly shown in the
Appendix.

\section{Perturbing  the metric of 5D Hard-Wall models}

\subsection{5D flat Hard Wall model}

The case of gauge vector field in a flat extra-dimension slice $0<
z< z_0$, \emph{i.e.} with warp factor $w(z)=1$,  has a nice
interpretation as the continuum limit of deconstruction models
\cite{Son:2003et}. Its simplicity allows for analytic calculations
of the first order perturbative corrections to masses and coupling
constants due to a metric perturbation. We report here the
corresponding formulae. Moreover, eq.(\ref{compact}) can be used
to extract the sub-leading terms in the expansion of the two-point
function for large Euclidean momenta $p^2=-Q^2$ which is also an
important effect of the metric deformation establishing an
holographic link between metric deformations of the 5D theory and
non-vanishing condensates of the 4D dual theory. In fact, in the
4D gauge gauge theory, sub-leading terms of the two-point function
in the deep Euclidean kinematical region are produced by
non-perturbative effects of condensates \cite{Shifman:1978bx}.

Turning to the case of flat 5D extra-dimension on a finite
interval, the normalized wave-functions are
\begin{equation}
\varphi_n^{(0)}(z)=\sqrt{\Frac{2}{z_0}}\sin\left(\kappa_{0,n}\frac{z}{z_0}
\right),\;\;{\rm with} \;\;\kappa_{0,n}=\frac{2n+1}{2}\pi,
\end{equation}
with eigenvalues
\begin{equation}
\lambda_n^{(0)}=-(m_n^{(0)})^2=-\left(\frac{\kappa_{0,n}}{z_0}\right)^2.
\end{equation}
The bulk-to-boundary propagator is the solution of (\ref{EOM}),
with  $w_0(z)=1$ and satisfying the boundary conditions
(\ref{BC0}):
\begin{equation}
\cV_0(z,p)=\left(\frac{\sin(pz_0)}{\cos(pz_0)}\sin(pz)+\cos(p
z)\right)
\end{equation}
Using (\ref{sigma0}), the two-point function can be written
\begin{equation}
\Sigma_0(p^2)=-\cV'_0(0,p)=-p\tan(p z_0).
\end{equation}
The bulk-to-boundary  can also be expanded in series of
eigenfunctions as follows
\begin{equation}
\cV_0(z,p)=-\frac{2}{z_0^2}\sum_{n=0}^\infty \,
\Frac{\kappa_{0,n}}{p^2-\left(\Frac{\kappa_{0,n}}{z_0}
\right)^2}\,\sin\left(\kappa_{0,n}\Frac{z}{z_0}\right).
\end{equation}
Then, the two point function admits the bound-states decomposition
\begin{equation}
\Sigma_0 (p^2)=\frac{2}{z_0^3}\sum_{n=0}^\infty
\Frac{\kappa_{0,n}^2}{p^2-\left(\Frac{\kappa_{0,n}}{z_0}\right)^2}.
\end{equation}
and one  can read the values of the unperturbed masses of the
vector resonances from the poles and their coupling constants from
the corresponding residues at the pole:
\begin{equation}
m_n^{(0)}=\Frac{\kappa_{0,n}}{z_0},
\;\;F_n^{(0)}=\sqrt{\Frac{2}{z_0^3}}\,\kappa_{0,n}.
\end{equation}
 We now use  the results of Secs. 3,  to evaluate the
the correction to the two-point function and the effects on
resonance masses and decay constants, due to a  deformation of the
flat metric of the form
\begin{equation}
 h(z)=\eta\,(z/z_0)^4.\label{pertz4}
\end{equation}
 As we
shall see in the  in the next section, this choice of the
deformation is suggested by the fact that,  in the AdS/CFT
correspondence, it would reproduce the  effects of a  4D operator
of conformal dimension four such as the gluon condensate. We
obviously assume in our perturbative approach  the dimensionless
constant $\eta$ to be  small, and consider only corrections of
first order in $\eta$. Notice that $h(0)=0$.

Inserting $\cV_0(z,p)$ above  into the expression (\ref{compact}),
the resulting integrals can be easily evaluated and one gets for
the first order correction of the two-point function
\begin{equation}
\Sigma_1(p^2)=-\Frac{\eta}{4z_0^4p^3}\left(\Frac{2pz_0(2p^2z_0^2-3)}{\cos^2(pz_0)}+
 6\tan(pz_0)\right)\label{Sigma1flat}
\end{equation}
It is convenient to use the identity $a+\eta \;b\sim a/(1-\eta\;
b/a)$, valid to first order in $\eta$, and rewrite  1st order
corrected two-point function in the form
\begin{equation}
\Sigma_0(p^2)+\Sigma_1(p^2)\sim
-\Frac{p\sin(pz_0)}{\cos(pz_0)-\eta\Frac{1}{4z_0^4p^4}
\left(\Frac{2pz_0(2p^2z_0^2-3)}{\sin(pz_0)}+
 6\cos(pz_0)\right)}\label{Sigma0and1flat}
\end{equation}
Perturbed masses and decay constants can be extracted from
(\ref{Sigma0and1flat}) by looking for the zeroes of the
denominator, \emph{i.e.} the positions of the poles, and then
extracting  the corresponding residues.

Explicitly, once that the two-point function has been put in the
form of a ratio of two functions $\Sigma(p^2)=\cN(p)/\cD(p)$ as
above, residues are evaluated by expanding, around the poles,
\emph{i.e.} around the zeros $m_n$ of the denominator $\cD(q)$,
and thus
\begin{equation}
F_n^2=2 m_n \Frac  {\cN(m_n)}{\cD'(m_n)}.
\end{equation}
 In a perturbative approach,
$\cD(p)=\cD_0(p)+\eta \,\cD_1(p)$ and,  to first order in $\eta$,
the following expressions for masses and residues  are obtained
\begin{equation}
m_n=m_n^{(0)}+\eta\,
m_n^{(1)}=m_n^{(0)}\left(1-\eta\,\Frac{\cD_1(m_n^{(0)})}{\cD'_0(m_n^{(0)})\,m_n^{(0)}}\right)
\end{equation}
and
\begin{equation}
F_n=F_n^{(0)}\left(1+\Frac{\eta}{2}\left(\Frac{m_n^{(1)}}{m_n^{(0)}}+
\Frac{\cN'_0(m_n^{(0)})}{\cN_0(m_n^{(0)})}m_n^{(1)}-
\left(\Frac{\cD_0'(m_n^{(0)})\cD_1'(m_n^{(0)})-\cD_0''(m_n^{(0)})\cD_1(m_n^{(0)})}
{\cD_0'(m_n^{(0)})^2} \right)\right)\right)
\end{equation}

In the case of 5D flat  slice, previous formulae boil down to:
\begin{equation}
m_n=m_n^{(0)}\left(1-\frac{\eta}{2}\frac{2\kappa_{0,n}^2-3}{\kappa_{0,n}^4}\right)
\end{equation}
\vskip .05cm
\begin{equation}
F_n=F_n^{(0)}\left(1-\frac{\eta}{2}\frac{1}{\kappa_{0,n}^2}\right)
\end{equation}
Corrections to masses and decay constants can also be obtained
using the perturbation theory for the bound-state eigenfunctions
in Sec.3. It is not difficult to check that the  corrections to
the masses coincide with those obtained from perturbation formula
(\ref{correlambda}) for the eigenvalues. We have also checked that
the corrections to the decay constants coincides numerically with
the ones obtained using perturbation formulae (\ref{correeigen})
and taking first derivatives of the perturbed eigenfunctions
evaluated at $z=0$ as in (\ref{pertdecay}).

Finally, using (\ref{Sigma1flat}) with Euclidean momentum
$p^2=-Q^2$ one has in the large-$Q^2$ that the two-point function
receives a subleading contribution from the metric deformation
given by
\begin{equation}
\Sigma(-Q^2)\approx Q\left(1+\Frac{3\eta}{2}\Frac{1}{z_0^4
Q^4}\right).
\end{equation}
\subsection{AdS$_5$  Hard-Wall model}

As explained in the Introduction, the case of gauge vector field
in a 5D AdS slice $0< z< z_0$, i.e. with warp factor $w(z)=1/z$,
is the one originally proposed as  holographic dual of QCD. There,
the AdS/CFT dictionary, relating 5D fields  and 4D operators  is
supposed to hold and it  gives a good understanding of what 4D
effect one is actually evaluating. Referring to the original
papers for the details, we limit ourself to repeat the same sort
of calculations we did in the previous section for the 5D flat
space, and concentrate again on the effects of a perturbation of
the form (\ref{pertz4}), whose link to 4D condensates can be given
a much solid ground in the AdS case \cite{Hirn:2005vk}.

Formulae become somewhat more complicated due to the appearance of
Bessel functions both in the expression of the unperturbed
bulk-to-boundary propagator and in the eigenfunctions. First order
correction is obtained in terms of  integrals containing  Bessel
functions, which, in principle, could be written in terms of
higher transcendental functions, but for practical purposes, can
be evaluated numerically, once the value of dimensional parameter
$z_0$, giving the size of the 5D slice and the 4D energy scale at
which 4D conformal invariance is broken, is fixed.

In the AdS case the  eigenfunction of the $n$-th bound-state is
given by
\begin{equation}
\varphi_n^{(0)}(z)=\Frac{\sqrt{2}}{|J_1\left(\gamma_{0,n}\right)|}\Frac{z}{z_0}
J_1(\gamma_{0,n}\Frac{z}{z_0}) ,\label{eigenAdS}
\end{equation}
with $\gamma_{0,n}$ being the $n$-th zero of the Bessel function
of order zero, \emph{i.e.} $J_0(\gamma_{0,n})=0$.

The  eigenvalues are
\begin{equation}
\lambda_n^{(0)}=-(m_n^{(0)})^2=-\left(\Frac{\gamma_{0,n}}{z_0}\right)^2.
\end{equation}
The bulk-to-boundary propagator is the solution of (\ref{EOM}),
with $w_0(z)=1/z$ and satisfying the boundary conditions
(\ref{BC0}) is:
\begin{equation}
\cV_0(p,z)=\Frac{z}{\epsilon}\Frac{Y_0(p z_0)J_1(p z)-J_0(p
z_0)Y_1(p z)} {Y_0(p z_0)J_1(p\epsilon)-J_0(p z_0)Y_1(p
\epsilon)},\label{V0AdS}
\end{equation}
There is a divergence as the UV cut-off $\epsilon$ is sent to
zero. A singular behavior is shown by the  two-point function:
\begin{equation}
\Sigma_0(p^2)\approx p^2 \log
\left(\Frac{p\epsilon}{2}\right)-\Frac{\pi p^2}{2} \Frac{Y_0( p
z_0)}{J_0(p z_0)}.\label{Sigma0AdS}
\end{equation}
The logarithmic behavior in the UV is actually welcome since it is
identified with the contribution of the parton loop at high
momenta. The finite part in (\ref{Sigma0AdS}) contains the
information on the resonance poles, which,  are located at the
zeroes of $J_0(p z_0)$. From the residues  one gets the decay
constants:
\begin{equation}
F_n^{(0) 2}=\Frac{\pi\gamma_{0,n}^3}{z_0^4}
\Frac{Y_0(\gamma_{0,n})}{J_1(\gamma_{0,n})}.\label{res0AdS}
\end{equation}
We can  explicitly show the bound-state expansion of the two-point
function by means of the Kneser-Sommerfeld formula
\cite{Grigoryan:2007vg} (valid for $ z \leq z_0 $)
\begin{equation}
\label{KS} \frac{Y_0(p z_0)J_0(p z) - J_0(p z_0)Y_0(p
z)}{J_0(pz_0)} \nonumber  = -\frac4\pi  \, \sum_{n = 1}^{\infty}
\frac{J_0(\gamma_{0,n}z/{z_0})} {[J_1(\gamma_{0,n})]^2(p^2z^2_0 -
\gamma^2_{0,n})} \ ,
\end{equation}
The limit  $ z \rightarrow 0 $  gives  a logarithmically divergent
series, which however correctly describes the behavior around the
poles:
\begin{equation}
\label{vecor} \Sigma_0(p^2) = \frac{2 p^2}{z^2_0}\sum_{n =
1}^{\infty} \Frac{[J_1(\gamma_{0,n})]^{-2}}{p^2
-\left(\Frac{\gamma_{0,n}}{z_0}\right)^2} ,
\end{equation}

From this expression one  reads the values of the (unperturbed)
masses of the vector  resonances from the poles and their coupling
constants from the corresponding residues at the pole. The
agreement with the value (\ref{res0AdS}), is provided by the
identity $\gamma_{0,n}J_1(\gamma_{0,n})Y_0(\gamma_{0,n})=2/\pi$.

 We now use  the results of Sec.3,  to calculate the effects
 of the   perturbation  of
a metric of the form (\ref{pertz4}) on the two-point function, and
on resonance masses and decay constants. This choice of the
deformation is suggested by the fact that,  in the AdS/CFT
correspondence, it would reproduce the effects of a  4D operator
of conformal dimension four such as the gluon condensate. We again
assume in our perturbative approach the dimensionless constant
$\eta$ to be a small, and consider only first order corrections in
$\eta$.

Inserting $\cV_0(z,p)$ of eq.(\ref{V0AdS}) into the expression
(\ref{compact}) one obtains integrals containing Bessel functions
and, taking the limit $\epsilon\rightarrow 0$, the first order
correction of the two-point function  can be written
\begin{equation}
\Sigma_1(p^2)=\eta\Frac{\pi^2}{p^2z_0^4}\left[\Frac{Y_0(p
z_0)^2}{J_0(p z_0)^2}\cI_{J_1J_0}(p z_0)-\Frac{Y_0(p z_0)}{J_0(p
z_0)}(\cI_{J_1Y_0}(p z_0)+\cI_{J_0Y_1}(p z_0))+\cI_{Y_0Y_1}(p
z_0)\right],\label{Sigma1AdS}
\end{equation}
where we defined the following integral of Bessel functions
$Z_n(x)$:
\begin{equation}
\cI_{Z_mZ_n}(x)\equiv\int_0^xZ_m(y)\,Z_n(y)\,y^4\,dy.
\end{equation}
Proceeding as in the flat case, we can extract perturbed masses
and decay constants from poles and  residues of the perturbed
two-point function. The final results are
\begin{equation}
m_n=m_n^{(0)}+\eta\, m_n^{(1)}=m_n^{(0)}\left(1-\eta\Frac{4
\cI_{J_1J_0}(\gamma_{0,n})}{\gamma_{0,n}^6 J_1(\gamma_{0,n})^2
}\right)\label{masscorrAdS}
\end{equation}
\vskip .05cm
\begin{eqnarray}
&F_n^2 =F_n^{(0)^2}\left[1+\eta\left(-\Frac{\pi}{\gamma_{0,n}^4}(
\cI_{J_0Y_1}(\gamma_{0,n})+\cI_{Y_0J_1}(\gamma_{0,n}))
-\Frac{1}{2}\Frac{m_n^{(1)}}{m_n^{(0)}}\right.\right.\label{decaycorrAdS}\\
&\left.\left.
+\Frac{\pi}{\gamma_{0,n}^4}\Frac{Y_1(\gamma_{0,n})}{J_1(\gamma_{0,n})}
\cI_{J_0J_1}(\gamma_{0,n})
+\Frac{1}{4}\Frac{m_n^{(1)}}{m_n^{(0)}}\gamma_{0,n}\Frac{J_2(\gamma_{0,n})}{J_1(\gamma_{0,n})}
-\Frac{1}{2}\Frac{m_n^{(1)}}{m_n^{(0)}}\gamma_{0,n}\Frac{Y_1(\gamma_{0,n})}{Y_0(\gamma_{0,n})}
\right) \right]\nonumber
\end{eqnarray}
As an independent check, we can apply the  perturbation theory of
Sec.3.2 to the AdS bound-states (\ref{eigenAdS}). Using
eq.(\ref{pertmass})  one recovers the same expression
(\ref{masscorrAdS}) for  mass corrections; through the
perturbation series (\ref{correeigen})
 and from eq.(\ref{pertdecay}), one finds
numerical agreement with the values of the corrections to the
decay constants one gets from (\ref{decaycorrAdS}).

In Table.1, these corrections are shown for the first few vector
and axial vector resonances. In order to get these numerical
values,  we had to restore in our formulae the 5D coupling
constant $g_5^2$ and fix it to $12\pi^2/N_c$ by matching the
logarithmic  term of the two-point function to the value of the
QCD parton loop. We have also posed $z_0=3.1 \times 10^{-3}\,{\rm
MeV}^{-1}$ in order to have that the mass of the first vector
resonance coincide with the value of the $\rho$ meson mass of
about $776\, \rm{MeV}$.

The values for the axial-vectors have been obtained working in the
HW model of  \cite{Hirn:2005nr}, where they are 5D gauge fields
satisfying the same 5D field equations  as the vectors, but
different IR boundary conditions at $z_0$, {\emph{i.e.}
$\varphi_n^A(z_0)=0$, in order to break chiral symmetry. This
leads to different unperturbed masses, decay constant,
wave-function and bulk-to-boundary propagator; however our
perturbative treatment of the correction due to the metric
deformation (\ref{pertz4}) can be done analogously to what we have
explicitly  illustrated in the case of vectors.
\begin{table}[tpb]%
\label{deltaMandF}\centering %
\begin{tabular}{|c|c|c|c|c|}
\hline
& & &  &  \\
Resonance  & $m_n^{(0)}$ (MeV) & $(\delta m_n/m_n^{(0)})/\eta$ &
$10^{-3}\;F_n^{(0)}$ (MeV$^2$)
 & $(\delta F_n/F_n^{(0)})/\eta$\\
& & &  &  \\
\hline\hline
$\rho^1$  & 776 & -0.15 & 109 & -0.23\\
\hline
$\rho^2$  & 1781 & -0.041 & 380 & -0.044\\
\hline
$\rho^3$  & 2792 & -0.017 & 747 & -0.018\\
\hline \hline
$a_1^1$  & 1236 & 0.045 & 223 & 0.049\\
\hline
$a_1^2$  & 2263 & 0.013 & 548 & 0.011\\
\hline
$a_1^3$  & 3282 & 0.006 & 955 & 0.009\\
\hline
\end{tabular}
{\caption{\small Corrections to vector and axial vector masses and
decay constants due to the  deformation $h(z)=\eta\, (z/z_0)^4$ of
the metric of a 5D AdS slice.} }
\end{table}

The mass corrections of the first vector and axial vector
resonance, \emph{i.e.} the $\rho$ and the $A_1$, agree with those
reported in Ref.\cite{Hirn:2005vk} once our parameter $\eta$ is
written as in terms of their $o_4$, both related to the gluon
condensate as follows:
\begin{equation}
\eta=\Frac{9\pi^2}{4} \ o_4 = \Frac{3\pi}{16 N_c}z_0^4\,\langle
\alpha_s  G_{\mu\nu}G^{\mu\nu}\rangle. \label{frometatoo4}
\end{equation}}
 We also notice that $\eta$ is almost equal to the
parameter $g$ of Ref.\cite{Bijnens:1992uz}, which expresses the
gluon condensate effects in the Extended Nambu-Jona Lasinio model
of low energy hadron phenomenology: one has $\eta=(9/8)(M_Q z_0)^4
g$, where $M_Q$ is the constituent quark mass of that model.

The relation between  $\eta$  and  the gluon condensate can be
established  by the fact that in presence of the metric
perturbation a non-vanishing sub-leading inverse power correction
appears in the two-point function at large Euclidean momentum.
Explicitly in this limit and using $p^2=-Q^2$  in (\ref{Sigma1AdS})   the two-point
function receives a sub-leading contribution from the metric
deformation  given by
\begin{equation}
\Sigma(-Q^2)\approx -\Frac{N_c}{24\pi^2}Q^2\left[\log
\left(\Frac{Q\epsilon}{2}\right)^2-\Frac{16\,\eta}{3 z_0^4
Q^4}\right].\label{2ptHW}
\end{equation}
The coefficient of the sub-leading term can then be  matched
with the one produced in QCD by a non-vanishing  gluon condensate:
\begin{equation}
  \Sigma \left( - Q^2 \right) = - \frac{N_c}{24 \pi^2} Q^2
  \left[\log\left(\frac{Q^2}{\mu^2}\right)
 - \frac{(\pi \alpha_s  \left\langle G_{\mu \nu}
  G^{\mu \nu} \right\rangle/N_c)}{Q^4}+ \mathcal{O} \left( \frac{1}{Q^6} \right) \right],
  \label{QCD2pointsub}
\end{equation}
From (\ref{2ptHW}) and (\ref{QCD2pointsub}) it follows that $\eta$
has the same sign of the gluon condensate, which experimentally
results to be positive (see, for instance, the up-to-date
discussion in \cite{Narison:2009vy} and references therein). The
use of the value, $\langle \alpha_s  G_{\mu\nu}G^{\mu\nu}\rangle
\sim 6.8\cdot 10^{-2}\ {\rm GeV}^4 $, consistent with the present
determination (see \cite{Espriu:1989ff}) for the gluon condensate,
leads to   $\eta\sim 1.26$ in eq.(\ref{frometatoo4}); however the
proper value to use for holographic models should only come from a
fit of the full holographic predictions to the data but,
anticipating the phenomenological discussion in the next
Section, we can say that it is close to a reasonable value.

 It should be noticed that, in the HW model
of \cite{Hirn:2005nr}, where chiral symmetry breaking is imposed
through different boundary conditions for vector and  axial vector
gauge fields, the deformation of the metric produces the same
sub-leading term at large Euclidean momentum both for the vector
and the axial-vector two-point functions. The effects of different
IR b.c. is indeed exponentially suppressed at large Euclidean
momentum. This is in agreement with the non-chiral nature of the
gluon condensate in QCD.

Let us  come to a consistency check of our results. In any HW
model, completeness property of the resonance wave-functions give
rise to sum rules involving vector and axial vector meson masses
and decay constant and other parameters of the low energy chiral
lagrangian. Explicitly, one has the following sum rules
\cite{Hirn:2005nr},\cite{Hirn:2007bb}

\begin{eqnarray}
\sum_n  g_{n}^2&=& 8\, L_1\label{Sum1}
\\
\sum_n f_{n} g_{n}&=&2L_9\label{Sum2}
\\
\sum_n  g_{n}^2 m_{n}^2&=&\Frac{f_{\pi}^2}{3}\label{Sum3},
\\
\sum_n f_{n} g_{n}m_{n}^2&=&f_{\pi}^2\label{Sum4},
\end{eqnarray}
where $L_1$ and $L_9$ are two of the Gasser-Leutwyler coefficients
of the $O(p^4)$ terms of the chiral Lagrangian \footnote{The HW
models of \cite{Hirn:2005nr} automatically enforce, in the
$SU_L(3)\times SU_R(3)$ case, the relations $L_3=6\,L_1=3\,L_2$
and $L_9+L_{10}=0$}, $f_{\pi}=92.3$ MeV is the pion decay
constant, $f_{n}=F_n/m_n^2$ and $g_{n}$ are the pion decay
constants and two-pion-one-vector couplings as they are usually
written in the Chiral Lagrangian formulation of spin-1 vector
field sector,
\begin{eqnarray}
{\cal L}_V \, &=& \, - \, \Frac{f_V}{2 \sqrt{2}} \langle \, V_{\mu
\nu} \, f_{+}^{\mu \nu} \, \rangle \, - \, \Frac{i g_V}{2
\sqrt{2}} \, \langle V_{\mu \nu} \, [ \, u^{\mu} \, , \, u^{\nu}
\, ] \rangle + \ldots \label{spinone}
\end{eqnarray}
where $V_{\mu\nu}$ is the $V_\mu$ field strength, $u_{\mu} \, = \,
i \, u^{\dagger} \, D_{\mu} \, U \, u^{\dagger}$, with $U=u^2$
being the chiral field and $f_{+}^{\mu \nu} \, = \, u \, F_L^{\mu
\nu} \, u^{\dagger} \, +\, u^{\dagger} \, F_R^{\mu \nu} \, u $,
with $F_{L,R}^{\mu \nu}$ external sources. For the lowest
resonance,  the $\rho$, one would have $f_V\equiv f_\rho\equiv
f_1$ and $g_V\equiv g_\rho\equiv g_1$. Our notations in
Eq.(\ref{spinone}) agree with those of Ref.\cite{Hirn:2005nr}.

Notice that in writing the sum rules (\ref{Sum1}-\ref{Sum4}) we
are assuming $\chi$SB and  the pion wave function of
 in the 5D holographic model of QCD of Ref.\cite{Hirn:2005nr}.  A
different approach is followed in Ref.\cite{Erlich:2005qh}.

The physical 4D meaning of the sum rules above is that they ensure
soft high energy behavior of some quantities relevant for physical
processes  such as the pion elastic scattering amplitude
(\ref{Sum1}), the vector form factor (\ref{Sum2},\ref{Sum4}). The
sum rule (\ref{Sum3}) is the analog of the KSFR for the infinite
tower of resonances one obtains in these HW model. Note that in
models with vector meson dominance, corresponding to just the
first resonance, the natural value KSFR is a  factor \lq2\rq   \
instead of  \lq 3\rq \  which is what one gets instead in these HW
models.

In the expression of the $L_i$ there enters the pion wave function
$\alpha(z)$ which, for a metric of the form (\ref{metric}) is
solution of the equation
\begin{equation}
\partial_z\left(w(z)\partial_z\alpha(z)
\right)=0,
\end{equation}
with boundary conditions
\begin{equation}
\alpha(0)=1,\;\; \alpha(z_0)=0.
\end{equation}
In presence of a deformation of the original AdS metric, the
function $\alpha(z)$ will also get corrections. At first order in
$\eta$ one has
\begin{equation}
\alpha(z)=1-\Frac{z^2}{z_0^2}-\Frac{\eta}{3}\left(\Frac{z^2}{z_0^2}-
\Frac{z^6}{z_0^6}\right)\label{pionwave}
\end{equation}

The  value of  $g_n$ is obtained as a 5D  overlapping integral
containing $\alpha(z)$ and the vector wave-function
$\varphi_n(z)$, \emph{i.e}:
\begin{equation}
g_n=\frac{\sqrt{2}}{2g_5}\int_0^{z_0}w(z)(1-\alpha(z)^2)\varphi_n(z)dz.
\end{equation}

As we may apply the perturbation formulae (\ref{correeigen}) to
get the correction to the vector wave functions $\phi_n(z)$, we
are able to evaluate numerically the corrections to the couplings
$g_n$'s. Their values for the firs few resonances are shown  in
the table and the plot of Fig. 1.
\begin{figure}[h]
\begin{minipage}[b]{6cm}
{
 \hspace{1cm}\vspace{.5cm}
\begin{tabular}{|c|c|c|}
\hline
 &  &  \\
Resonance  & $g_{n}^{(0)}$  & $(\delta
g_{n}/g_{n}^{(0)})/\eta$ \\
 &  &  \\
\hline\hline
$\rho^1$  & 0.0538 & 0.34 \\
\hline
$\rho^2$  & -0.0019 & -2.31 \\
\hline
$\rho^3$  & 0.0003 & -3.72\\
\hline
\end{tabular}
}
\end{minipage}
\hskip3cm
\begin{minipage}[b]{8cm}
\includegraphics[width=7cm,height=4cm]{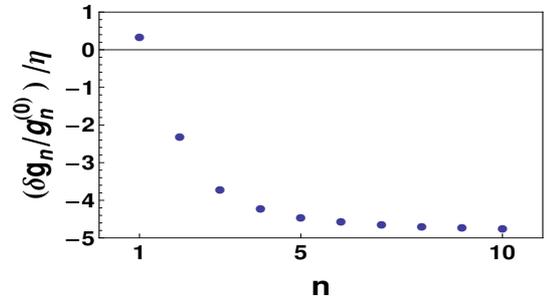}
 \end{minipage}
{\caption{\small Corrections to the one-vector-two-pions couplings
for the first few vector resonances  due to the deformation
$h(z)=\eta\, (z/z_0)^4$ of the metric of a 5D AdS slice.} }
 \end{figure}

We now report the expressions of the Gasser-Leutwyler coefficients
of the chiral lagrangian obtained in the AdS HW model
\cite{Hirn:2005nr}, at first order in the metric perturbation of
eq.(\ref{pertz4}). Using $1/g_5^2=N_c/12\pi^2$, one has:

\begin{eqnarray}
f_\pi&=&\sqrt{\Frac{1}{g_5^2}\int_0^{z_0}w(z)(\alpha'(z))^2 dz}=
\Frac{\sqrt{N_c}}{\sqrt{6}\pi
z_0}\left(1+\Frac{1}{6}\eta\right)\label{fpipert}
\\
L_1&=&\Frac{1}{32 g_5^2}\int_0^{z_0}w(z)(1-\alpha(z)^2
)^2dz=\Frac{11
N_c}{9216\pi^2}\left(1+\Frac{2}{3}\eta\right)\label{L1pert}
\\
L_9&=&\Frac{1}{4 g_5^2}\int_0^{z_0}w(z)(1-\alpha(z)^2)dz=\Frac{
N_c}{64\pi^2}\left(1+\Frac{25}{54}\eta\right)\label{L9pert}
\\
L_2&=&2L_1,\;L_3=-6L_1,\;L_{10}=-L_9. \label{L10pert}
\end{eqnarray}

Notice that the smallness of the numerical coefficients in front
of $\eta$ in (\ref{fpipert}-\ref{L9pert}) makes these corrections
perturbative even for $\eta\sim O(1)$. When the perturbation
parameter $\eta$ is written in terms of $o_4$, following
(\ref{frometatoo4}), our  expressions (\ref{fpipert}-\ref{L9pert})
coincide with the ones in Ref.\cite{Hirn:2005vk}.

An important result of holographic models is that the sum rules
(\ref{Sum1}-\ref{Sum4}) still hold with the perturbed values of
the $L_i$'s, of $f_\pi$ and of the single resonances masses $m_n$,
decay constants $f_n$ and coupling to two pions $g_n$. As such, we
can also  address an issue similar to the one raised in
Ref.\cite{Chivukula:2004kg} in the context of deconstruction
models with unperturbed metrics. The authors  wondered how fast is
the convergence of the sum over the resonances  in the first and
second Weinberg sum rule (WSR): these sums, compared to eqs.
(\ref{Sum1}-\ref{Sum4})   involve also axial vectors. Their
findings was that actually a large number of terms  was needed;
for instance in the case of deconstruction with \lq\lq $\cosh z
$\rq\rq background, the one leading to AdS in the continuum limit,
and twenty lattice sites, practically all resonances had to be
considered in order to get a small deviation in  the WSR.

Authors of Ref. \cite{Becciolini:2009fu} address a  different
question in  QCD  deconstruction: how many sites are needed to
have a  a good phenomenological picture. They find that already a
model with three or four sites reproduces the relevant features of
holographic models of QCD.

These two issues led us to investigate how good is the convergence
of the sum rules in a perturbed metrics  in Eqs.
(\ref{Sum1}-\ref{Sum4}), i.e. how many resonances are needed to
achieve a good descriptions of Eqs. (\ref{fpipert}-\ref{L9pert}).
This  is even more reasonable, if some pattern  of low resonance
dominance shows up in the theory. Indeed, for all sum rules the
convergence remains quite good even after the corrected values of
$m_n$, $f_n$ and $g_n$ are inserted, and what is more important,
the numerical values converge rapidly with just few terms in the
sum to those obtained from the perturbations of $f_{\pi}$, $L_1$
and $L_9$, what gives us an important check. Actually, a dominance
of the few lowest resonances appear, as it is illustrated in Fig.
2, where saturation of each sum rule by the lowest resonances is
shown in both the unperturbed and in the first order corrected
case.

\begin{figure}[!t]\centering \epsfysize=14cm\epsfxsize=14cm%
\epsfbox{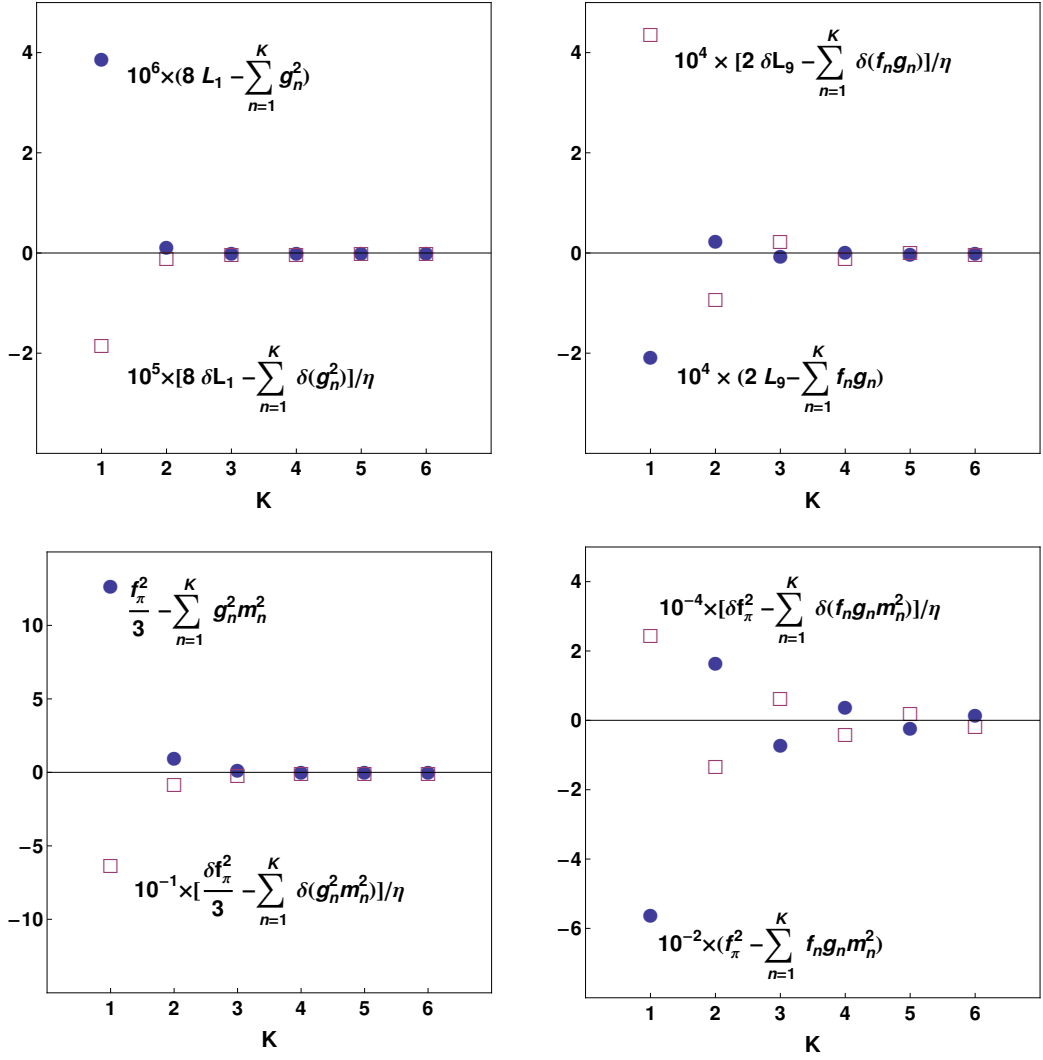} {\caption{\small The four plots show the
result of adding the first $K$ terms in the two sum rules
(\ref{Sum1}-\ref{Sum4}) in the HW model of Ref.
\cite{Hirn:2005nr}, both in the case of pure AdS metric (dots) and
in presence of first order corrections in the  metric deformation
(\ref{pertz4}) (squares). To put both in the same plot, the
differences in the unperturbed and in the perturbed case have been
magnified by suitable factors, as indicated. One easily recognizes
that the sum rules are dominated by the contribution of the first
few resonances also in presence of the metric perturbation. In the
plots, the perturbative corrections
 in eqs. (\ref{fpipert}),(\ref{L1pert}) and (\ref{L9pert}) and
 for $m_n$, $f_n$ and $g_n$  (re-scaled by $\eta$) have been denoted
 by $\delta f_{\pi}$, $\delta L_1$ and $\delta L_9$ etc.. }}\label{psiv}
\end{figure}
\newpage
\section{Conclusions}

Having evaluated in eqs.(\ref{fpipert}-\ref{L9pert}), the gluonic
contributions to the  $L_i$'s and to  $f_\pi$ it is  interesting to check if they lead to
a better agreement with the known  experimental values.
Since
 we have been able to evaluate the effects of the
gluonic condensate on masses, decay constants and
vector-to-two-pion couplings of  single vector resonances we can
add  also these parameter to the discussion.

Before embarking in the numerical analysis, we  should stress
however  that the model we are considering has intrinsic
approximations that cannot be underestimated when trying to
compare it with the real dynamics of low energy QCD. As in any
Large-$N_c$ model of QCD we are neglecting order ${\cal O}(1/N_c)$ effects, such as $\pi\pi$-loops. We
also know that the model does not reproduce the Regge behavior. The same
mechanism of chiral symmetry breaking is strongly model dependent.
We have followed the one
proposed in Ref.\cite{Hirn:2005nr}, where  the pion is introduced
as a 5D \emph{non-local} object linked to a Wilson line between
the two boundary of the $AdS_5$-slice.
This leads to the
expression (\ref{pionwave}) for the pion wave function.
 In Refs.
\cite{Erlich:2005qh}, although the description of vector
resonances is the same, the spontaneous chiral symmetry breaking
is triggered by a 5D scalar field, dual to the $\bar{q}q$ operator
and not by IR boundary conditions.
We shall try to take care of these caveats in the numerical
discussion that follows.
As  in Ref.\cite{Hirn:2005nr} we work in the chiral limit.

We want to  distinguish between the predictions of the model for
those observables which involve the spectrum of resonance as a
whole as in the case for the Gasser-Leutwyler coefficients and the
pion decay constant and the predictions for parameters related to
a single resonance. Moreover, motivated by the previous
discussion, we shall estimate the systematic error of the model
for the prediction of the pion decay constant and of single
resonance parameters to be of about $10-20\%$. The numerical
analysis which we have performed gives indication in favor of
this approach.

We refer to Table 2 and Figure 3 to illustrate the results of the
analysis:   five Gasser-Leutwyler
coefficients, the pion decay constant and the mass, the decay
constant and the coupling to two pions of the lowest vector
resonance, identified with the $\rho$ meson, are the observables
that we are using in different fits.

\begin{table}[!ht]%
\label{GLcoeff}\centering %
\begin{tabular}{|c|c|c|c|c|}
\hline
 & & & &  \\
   &Vanishing gluon  & Fit of $L_i$ &
   Fit of $L_i$,  $f_\pi (20\%)$&
      \\
  & condensate&  and $f_\pi (20\%)$ & and ($m_{\rho}$, $f_{\rho}$, $g_{\rho}$)$(10\%)$
           & Experiment \\
  &$\eta=0, \, z_0(m_{\rho}^{phys})$ & $(\chi^2/ {\rm d.o.f.})=(3.2/4)$ &  $(\chi^2/{\rm d.o.f.})=(8.0/ 7)$&  \\
    & & & &  \\
      \hline\hline
    $ 10^3 L_1 $&$\  0.36 $& $\ 0.54$ & $\ 0.58$& $\ \ 0.4 \pm 0.3$\\
     \hline
    $ 10^3 L_2 $&$\ 0.72 $&$\ 1.08$& $\ 1.16$ & $\ 1.35 \pm 0.3$\\
     \hline
    $ 10^3 L_3   $&$ - 2.2  $& $ -3.2$ & $ -3.5$& $ - 3.5 \pm 1.1$\\
     \hline
    $ 10^3 L_9   $&$\ \ 4.7 $ & $\ \ 6.3 $& $\ \ 6.7 $& $\ \ \ 6.9 \pm 0.7$\\
     \hline
    $ 10^3 L_{10}  $&$ -4.7$ & $ -6.3 $& $ -6.7 $&$\ - 5.5 \pm 0.7$\\
     \hline
         $ f_{\pi} \, \text{(MeV)}  $&$\ 72.8$ & $\ 92.4 $& $\ 95.6 $&$\ 92.3 \pm 0.3$\\
     \hline
     $m_{\rho} \, \text{(MeV)}  $ & $\ 776^{\,\sharp}$ &  $\ 783^{\,\dag} $& $\ 769$&$\ 775.8 \pm 0.5$\\
     \hline
         $ f_{\rho}   $&$\ 0.18$ & $\ 0.19^{\,\dag} $& $\ 0.19 $&$\ 0.20 \pm 0.02^*$\\
     \hline
             $ g_{\rho}   $&$\ 0.054$ & $\ 0.067^{\,\dag} $& $\ 0.070 $&$\ 0.087 \pm 0.009^*$\\
     \hline

\end{tabular}
{\caption{\small Theoretical predictions and experimental values
for the five Gasser-Leutwyler coefficients $L_1,\, L_2, L_3, L_9,
L_{10}$ , the pion decay constant $f_\pi$ and the mass $m_\rho$,
the decay constant $f_\rho$ and the coupling $g_\rho$ of $\rho$ to
two pions. The first column refers to the values obtained in the
absence of gluon condensate, $\eta=0$, with $N_c=3$,  and with IR
scale of the model $z_0$ fixed by using the physical value
$m_\rho=776$ MeV (the $^\sharp$ denotes that there $m_\rho$ is an
input parameter). The second and third column show the result of
fitting the experimental values of the $L_i$, with their
experimental values, together with  $f_\pi$ (third column) and
together also with $m_\rho$, $f_\rho$ and $g_\rho$ (The values in
the second column have a $^{\,\dag}$ to remember that they have
been evaluated after the fit of the other observable had been
done). As explained in the text, to $f_\pi$ and to the physical
quantities of the $\rho$ meson have been assigned theoretical
errors of $20\%$ and $10\%$ respectively. The $^*$ on the values
of $f_\rho$ and $g_\rho$ in the last column indicate that errors
have been estimated in the Chiral Lagrangian theoretical
framework.}}
\end{table}

As benchmarks, in the  first column of Table 2 are shown
 the values that one obtains in the absence of
gluon condensate,  $\eta=0$,  with $N_c=3$, \emph{i.e.} correct
matching in the UV of the  vector two-point function with the QCD
parton loop, and with IR scale of the model $z_0$ fixed by using
the physical value $m_\rho=776$ MeV as input. The values of the
$L_i$'s appear to be underestimated.

As an attempt to improve the  predictions of the model, in absence
of gluon condensate, one could abandon the UV matching and let
$N_c$ vary. This was done in Ref. \cite{Hirn:2005nr}, where $N_c$
was fixed by using the value of $f_\pi$ as input, with $z_0$ still
fixed by the physical value of $m_\rho$. There resulted an
improved agreement with experimental values, at the cost of having
$N_c=4.3$, signalling that important physical effects were still
not captured by the holographic model without the condensate.

Let us now discuss the inclusion of the gluon condensate. As we
explained before, we shall focus on the prediction of the model
for observables which involve the complete spectrum of resonance
as is the case for the Gasser-Leutwyler coefficients and the pion
decay constant. In the second column of Table 2, we report the
values obtained by the fit of the $L_i$ and  $f_\pi$, using the
expressions (\ref{fpipert}-\ref{L10pert}), with an estimated
theoretical error on $f_\pi$ of $20\%$.  The values of $\eta$ and
$z_0$ which minimize the $\chi^2$ are $\eta=0.73$ and
$z_0=2.7\times 10^{-3}$ MeV$^{-1}$ (not far from the value
$3.1\times 10^{-3}$ MeV$^{-1}$ which would be obtained by fitting
$m_\rho$ without condensate), with $\chi^2$/{\rm d.o.f.}=3.2/4.
 The values obtained for $\eta$ and $z_0$ have been used to
compute the values of $m_\rho$, $f_\rho$ and $g_\rho$. The
$1\sigma$ region for $(z_0,\eta)$ is light shaded in the first
contour plot of Fig.3.

The third column of Table 2 shows the values obtained by adding
also the values of the physical parameters of the lowest vector
resonance, the $\rho$, with an estimated theoretical error of
$10\%$. The values of $\eta$ and $z_0$ which minimize the $\chi^2$
are $\eta=0.89$ and $z_0=2.7\times 10^{-3}$ MeV$^{-1}$, with
$\chi^2$/{\rm d.o.f.}=8.0/ 7. The $1\sigma$ region for
$(z_0,\eta)$ is light shaded in the second contour plot of Fig.3.
The determination of the best-$\chi^2$ values is sharpened at the
cost of having an higher $\chi^2$ per d.o.f.

\begin{figure}[!ht]\centering \epsfysize=8cm\epsfxsize=16cm%
\epsfbox{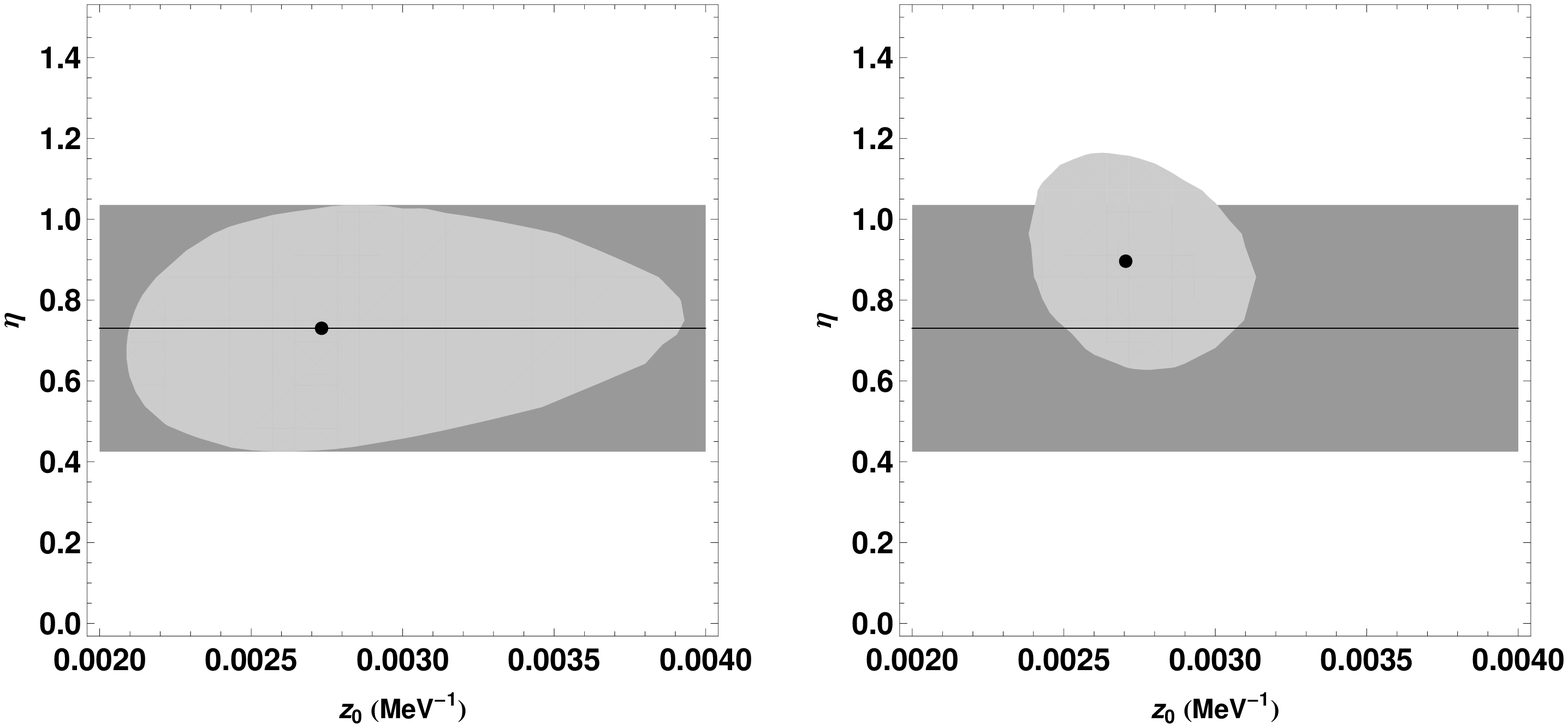} {\caption{\small Central values and
 $1\sigma$ regions for the values of the IR scale $z_0$
 and the condensate parameter $\eta$ are represented by a black dot and a light shaded
 region respectively  for the two fits  in Table 2. The left plot, with central
 values $\eta=0.73$ and $z_0=2.7\times 10^{-3}$ MeV$^{-1}$,
  corresponds  to the second column in Table 2: fit of the $L_i$'s and
  $f_\pi$. The right plot, with central values $\eta=0.89$ and $z_0=2.7\times 10^{-3}$ MeV$^{-1}$,
  corresponds to third column in Table 2: fit of the $L_i$'s, $f_\pi$,
  $m_\rho$, $f_\rho$ and $g_\rho$.
  The dark shaded band in both plots is the $1\sigma$ region for $\eta$ obtained
  by a fit of $L_i$'s values only, with the central value
  $\eta=0.73$ shown by the horizontal line.}}
\end{figure}

Using Eq.(\ref{frometatoo4}) one can derive from $\eta=0.89$ and
$z_0=2.7\times 10^{-3}$ MeV$^{-1}$ a prediction for the condensate
$\langle \alpha_s
G_{\mu\nu}G^{\mu\nu}\rangle=(8.7^{\,+9.4}_{\,-4.2})\times
10^{-2}\,{\rm GeV}^4$ which is compatible with the average
experimental value $\langle \alpha_s
G_{\mu\nu}G^{\mu\nu}\rangle_{\rm exp.}=(6.8\pm 1.13)\times
10^{-2}\,{\rm GeV}^4$ reported in Ref.\cite{Narison:2009vy}.

We believe that this simple analysis already shows that gluon
condensate corrections evaluated through metric deformation in HW
models leads to results qualitatively in agreement to what we
expect from QCD. We notice also that the trends of the numerical
 values of the Gasser-Leutwyler coefficients, that we have found,
in the presence of gluon condensate, and also those of vector and
axial vector masses and decay constants agree with those evaluated
in the ENJL model in \cite{Bijnens:1992uz}.

Now we want to observe another hint towards the connection between
the Chiral Quark Model \cite{Manohar:1983md,Espriu:1989ff} and
the AdS Hard-Wall model. As remarked in Ref.
\cite{deRafael:1995zv},  it just happens in the Chiral Quark Model
that, with neglect of the gluonic couplings,  for some of the
${\cal O}(p^4)$ terms in the QCD effective action ($L_1,\ L_2,\
L_3, \  L_9,\ L_{10}$) one gets finite results in the UV and IR
domains, {\it i.e.} cut-off independent values; of course this
does not hold once the  gluonic corrections are turned on. The
same is happening for the same $L_i$'s in the AdS HW model: no
dependence on $z_0$. Here we want to stress that this gives
further  support to the AdS picture vs the flat metric where the
same $L_i$'s show an explicit dependence on the IR cut-off, $z_0$.

In this paper we have  considered  the holographic description of
the effects of a gluon condensate in terms of a suitable
deformation of the  metric of a 5D Hard Wall model.  We have not
considered the effects of the $\langle \bar q q\rangle$ condensate
since our focus was on low energy QCD physics, once $\chi$SB has
been already triggered by some mechanism. Since the estimated
value for the gluon condensate is phenomenologically acceptable,
our treatment seems appropriate.  Moreover, the quark condensate
should not  be considered  in the comparison  with the Chiral
Quark model.

Comparing with previous literature, we have computed for the first
time  the corrections to the resonance decay constants, $f_n$ and
vector-to two-pion couplings, $g_n$ and we have shown the validity
of the sum rules even in presence of gluon condensate: this has
been the core of our calculation. We have also discussed the case
of flat metric. We have found substantial evidence  toward an
analogy of the AdS HW holographic model and the Chiral Quark
model. We have also performed a novel numerical analysis, which
leads to a satisfactory value for the gluon condensate, showing
that the model is able to capture the relevant physics properties.

\section*{Acknowledgements}

We thank Fabio Ambrosino,  Domenico Cuozzo and Guglielmo De Nardo
for discussions and help. This work has been supported in part by
the European Commission (EC) RTN FLAVIAnet under Contract No.
MRTN-CT-2006-035482 and by the Italian Ministry of Education and
Research (MIUR), project 2005-023102.

\appendix

\section{Appendix}

In this Appendix we give a  proof of the equivalence of the two
perturbative methods  developed in  Sec.3. We neglect the issue of
the convergence of the bound-state expansion (\ref{boundstates}).

 Let
us first show the useful identity
\begin{equation}
\sum_n\sum_m\Frac{w_0(0)^2\varphi_{ n}^{(0)'}(0)\varphi_{
m}^{(0)'}(0)}{p^2-(m_n^{(0)})^2}<\varphi_{ n}^{(0)}, \varphi_{
m}^{(0)}h>_0=0 \label{identity}
\end{equation}
The LHS of the previous expression can be written
\begin{equation}
\sum_n\Frac{w_0(0)^2\varphi_{
n}^{(0)'}(0)}{p^2-(m_n^{(0)})^2}<\varphi_{
n}^{(0)},\sum_m\varphi_{ m}^{(0)'}(0)
 \varphi_{ m}^{(0)}h>_0
 \end{equation}
Then,
\begin{eqnarray}
<\varphi_{ n}^{(0)},\sum_m\varphi_{ m}^{(0)'}(0)
 \varphi_{ m}^{(0)}h>_0=
\int_0^{z_0}\varphi_{ n}^{(0)}(z)\sum_m w_0(z)\varphi_{
m}^{(0)'}(0)\varphi_{ m}^{(0)}(z)h(z)dz=\nonumber\\
\Frac{d}{dz'}\left(\int_0^{z_0}\varphi_{ n}^{(0)}(z)\sum_m
w_0(z)\varphi_{ m}^{(0)}(z')\varphi_{
m}^{(0)}(z)h(z)dz\right)_{z'=0}= \Frac{d}{dz'}\left(\varphi_{
m}^{(0)}(z')h(z')\right)_{z'=0}=0
\end{eqnarray}
where  we used  the completeness relation (\ref{completeness}) for
the unperturbed eigenfunctions, and the last equality is a
consequence of the boundary condition (\ref{bceigen}) and of the
vanishing  of  $h(z)$ at the origin.

Using the identity (\ref{identity}) one can show, for instance,
that the two expressions which are obtained inserting the
bound-state expansion for $\cV(p,z)$  in the two different ways of
expressing the first order correction to the two-point function in
(\ref{compact}), actually coincide and give
\begin{equation}
-<\cV_0, D_1\cV_0>_0=-\sum_n\sum_m\Frac{w_0(0)^2\varphi_{
n}^{(0)'}(0)\varphi_{
m}^{(0)'}(0)}{(p^2-(m_n^{(0)})^2)(p^2-(m_m^{(0)})^2)}<\varphi_{
n}^{(0)}, D_1 \varphi_{ m}^{(0)}>_0. \label{correbound}
\end{equation}

This expression has to be identical to the one which we obtain
inserting in the bound-state expansion (\ref{boundstates}) the
perturbative corrections (\ref{correlambda}) and
(\ref{correeigen}) to masses and eigenfunctions, {\emph i.e.}
\begin{eqnarray}
\Sigma(p^2)=\sum_n\Frac{(w(0)\varphi_{
n}^{'}(0))^2}{p^2-m_n^2}=\sum_n\Frac{(w_0(0)(1+h(0))(\varphi_{
n}^{(0)'}(0)+\varphi_{
n}^{(1)'}(0)))^2}{p^2-((m_n^{(0)})^2+(m_n^{(1)})^2)}
\end{eqnarray}
Using (\ref{correlambda}) and (\ref{correeigen}), one has
\begin{equation}
\Frac{1}{p^2-((m_n^{(0)})^2+(m_n^{(1)})^2)}=\Frac{1}{p^2-(m_n^{(0)})^2}
\left(1+\Frac{(m_n^{(1)})^2}{p^2-(m_n^{(0)})^2}\right)=\Frac{1}{p^2-(m_n^{(0)})^2}
\left(1-\Frac{<\varphi_n^{(0)}, D_1\varphi_n^{(0)}>_0
}{p^2-(m_n^{(0)})^2}\right)
\end{equation}
and then
\begin{eqnarray}
\Sigma(p^2)=\Sigma_0(p^2)-\sum_n\Frac{(w_0(0)\varphi_{
n}^{(0)'}(0))^2}{(p^2-(m_n^{(0)})^2)^2} <\varphi_n^{(0)},
D_1\;\varphi_n^{(0)}>_0+\nonumber\\
2 \sum_n\Frac{w_0(0)^2\varphi_{
n}^{(0)'}(0)}{p^2-(m_n^{(0)})^2}\left(-\Frac{1}{2}<\varphi_n^{(0)},
\varphi_n^{(0)}h>_0\varphi_{ n}^{(0)'}(0)-\sum_{m\neq
n}\Frac{<\varphi_m^{(0)},
D_1\varphi_n^{(0)}>_0}{(m_n^{(0)})^2-(m_m^{(0)})^2}\varphi_{
m}^{(0)'}(0)\right)\label{lastSigma}
\end{eqnarray}
The next step is to notice that, as
\begin{equation}
\Frac{1}{p^2-(m_n^{(0)})^2}\;\Frac{1}{p^2-(m_m^{(0)})^2}=\Frac{1}{(m_n^{(0)})^2-(m_m^{(0)})^2}
\left(\Frac{1}{p^2-(m_n^{(0)})^2}-\Frac{1}{p^2-(m_m^{(0)})^2}\right),
\label{identity2}
\end{equation}
it is convenient to express the matrix elements $<\varphi_m^{(0)},
D_1\varphi_n^{(0)}>_0$ as the sum of the symmetric and
antisymmetric parts, which we denote by $S_{mn}=S_{nm}$ and
$A_{mn}=-A_{nm}$ respectively. With a little effort, one can show
that
\begin{eqnarray}
&S_{mn}=\Frac{1}{2}\left((m_n^{(0)})^2+(m_m^{(0)})^2\right)<\varphi_m^{(0)},\varphi_n^{(0)}
h>_0\,-\, <\varphi_m^{(0)'},\varphi_n^{(0)'} h>_0,&\nonumber\\
&A_{mn}=\Frac{1}{2}\left((m_n^{(0)})^2-(m_m^{(0)})^2\right)<\varphi_m^{(0)},\varphi_n^{(0)}
h>_0. &\nonumber
\end{eqnarray}
Then, the first order correction to $\Sigma(p^2)$ in
(\ref{lastSigma}) splits into four terms as follows:
\begin{eqnarray}
&-\displaystyle\sum_n\Frac{(w_0(0)\varphi_{
n}^{(0)'}(0))^2}{(p^2-(m_n^{(0)})^2)^2}<\varphi_n^{(0)},
D_1\varphi_n^{(0)}>_0 -\sum_n\Frac{(w_0(0)^2\varphi_{
n}^{(0)'}(0))^2}{p^2-(m_n^{(0)})^2}<\varphi_m^{(0)},\varphi_n^{(0)}
h>_0&\nonumber\\
&-2 \displaystyle\sum_n\displaystyle\sum_{m\neq
n}\Frac{w_0(0)^2\varphi_{m}^{(0)'}(0)\varphi_{
n}^{(0)'}(0)}{(p^2-(m_n^{(0)})^2)((m_n^{(0)})^2-(m_n^{(0)})^2)}S_{mn}&\label{big}\\
&- \displaystyle\sum_n\displaystyle\sum_{m\neq
n}\Frac{w_0(0)^2\varphi_{m}^{(0)'}(0)\varphi_{
n}^{(0)'}(0)}{p^2-(m_n^{(0)})^2}<\varphi_m^{(0)},\varphi_n^{(0)}
h>_0&\nonumber
\end{eqnarray}
Using the identity (\ref{identity}),  it can be shown that the sum
of the second and the fourth term in (\ref{big}) is zero. Using
the symmetry of $S_{mn}$ and the identity (\ref{identity2}), the
third term in (\ref{big}) can be rewritten
\begin{equation}
-\sum_n\sum_{m\neq n}\Frac{w_0(0)^2\varphi_{n}^{(0)'}(0)\varphi_{
m}^{(0)'}(0)}{(p^2-(m_n^{(0)})^2)(p^2-(m_m^{(0)})^2)}<\varphi_{
n}^{(0)}, D_1 \varphi_{ m}^{(0)}>_0, \label{nondiag}
\end{equation}

The first term in (\ref{big}) provides the diagonal terms not
present in (\ref{nondiag}) and allows to fully recover, as
promised, the expression (\ref{correbound}).


\begin{thebibliography}{999}
\bibitem{Maldacena:1997re}
  J.~M.~Maldacena,
  Adv.\ Theor.\ Math.\ Phys.\  {\bf 2}, 231 (1998)
  [Int.\ J.\ Theor.\ Phys.\  {\bf 38}, 1113 (1999)];
  S.~S.~Gubser, I.~R.~Klebanov and A.~M.~Polyakov,
  Phys.\ Lett.\ B {\bf 428}, 105 (1998);
  E.~Witten,
  Adv.\ Theor.\ Math.\ Phys.\  {\bf 2}, 253 (1998)

\bibitem{Sakai:2004cn}
  T.~Sakai and S.~Sugimoto,
  Prog.\ Theor.\ Phys.\  {\bf 113}, 843 (2005);

\bibitem{Erlich:2005qh}
  J.~Erlich, E.~Katz, D.~T.~Son and M.~A.~Stephanov,
  Phys.\ Rev.\ Lett.\  {\bf 95}, 261602 (2005);
  L.~Da Rold and A.~Pomarol,
  Nucl.\ Phys.\ B {\bf 721}, 79 (2005).

\bibitem{Hirn:2005nr}
  J.~Hirn and V.~Sanz,
  JHEP {\bf 0512}, 030 (2005);

\bibitem{Karch:2006pv}
  A.~Karch, E.~Katz, D.~T.~Son and M.~A.~Stephanov,
  Phys.\ Rev.\  D {\bf 74}, 015005 (2006)
  [arXiv:hep-ph/0602229].

\bibitem{Manohar:1983md}
  A.~Manohar and H.~Georgi,
  Nucl.\ Phys.\  B {\bf 234}, 189 (1984).

\bibitem{Espriu:1989ff}
  D.~Espriu, E.~de Rafael and J.~Taron,
  Nucl.\ Phys.\  B {\bf 345}, 22 (1990)
  [Erratum-ibid.\  B {\bf 355}, 278 (1991)].

\bibitem{Klebanov:1999tb}
  I.~R.~Klebanov and E.~Witten,
  Nucl.\ Phys.\  B {\bf 556}, 89 (1999)
  [arXiv:hep-th/9905104].


\bibitem{Erlich:2006hq}
  J.~Erlich, G.~D.~Kribs and I.~Low,
  Phys.\ Rev.\  D {\bf 73}, 096001 (2006)
  [arXiv:hep-th/0602110].

\bibitem{Hirn:2005vk}
  J.~Hirn, N.~Rius and V.~Sanz,
  Phys.\ Rev.\ D {\bf 73}, 085005 (2006)

\bibitem{Pomarol:1999ad}
  A.~Pomarol,
  Phys.\ Lett.\  B {\bf 486}, 153 (2000)
  [arXiv:hep-ph/9911294].

\bibitem{ArkaniHamed:2000ds}
  N.~Arkani-Hamed, M.~Porrati and L.~Randall,
  JHEP {\bf 0108}, 017 (2001)
  [arXiv:hep-th/0012148].

%
\bibitem{Son:2003et}
  D.~T.~Son and M.~A.~Stephanov,
  Phys.\ Rev.\  D {\bf 69}, 065020 (2004)
  [arXiv:hep-ph/0304182].

\bibitem{Shifman:1978bx}
  M.~A.~Shifman, A.~I.~Vainshtein and V.~I.~Zakharov,
  Nucl.\ Phys.\  B {\bf 147}, 385 (1979);
  Nucl.\ Phys.\  B {\bf 147}, 448 (1979).



\bibitem{Bijnens:1992uz}
  J.~Bijnens, C.~Bruno and E.~de Rafael,
  Nucl.\ Phys.\  B {\bf 390}, 501 (1993)
  [arXiv:hep-ph/9206236].

\bibitem{Narison:2009vy}
  S.~Narison,
  Phys.\ Lett.\  B {\bf 673}, 30 (2009)
  [arXiv:0901.3823 [hep-ph]].

\bibitem{Grigoryan:2007vg}
  H.~R.~Grigoryan and A.~V.~Radyushkin,
  Phys.\ Lett.\  B {\bf 650}, 421 (2007)
  [arXiv:hep-ph/0703069].




\bibitem{Hirn:2007bb}
  J.~Hirn and V.~Sanz,
  Phys.\ Rev.\  D {\bf 76}, 044022 (2007)
  [arXiv:hep-ph/0702005].


\bibitem{Chivukula:2004kg}
  R.~S.~Chivukula, M.~Kurachi and M.~Tanabashi,
  JHEP {\bf 0406}, 004 (2004)
  [arXiv:hep-ph/0403112].


\bibitem{Becciolini:2009fu}
  D.~Becciolini, M.~Redi and A.~Wulzer,
  JHEP {\bf 1001}, 074 (2010)
  [arXiv:0906.4562 [hep-ph]].




\bibitem{deRafael:1995zv}
  E.~de Rafael,
  arXiv:hep-ph/9502254.




\end{thebibliography}
\end{document}